\documentclass[10pt,a4paper]{article}
\usepackage[latin2]{inputenc}
\usepackage{amsmath}
\usepackage{amsfonts}
\usepackage{amssymb}
\usepackage{graphicx}
\usepackage{slashed}
\usepackage[a4paper,bindingoffset=0.2in,
            left=0.75 in,right=0.75 in,top=0.75 in,bottom=0.75  in,
            footskip=.25in]{geometry}
\usepackage{physics}

\begin{document}

\title{\textbf{Quantum mechanics with space-time noncommutativity}}
\author{Partha Nandi\footnote{S. N. Bose National Centre for Basic Sciences, Salt Lake, Kolkata 700106, India; Email: parthanandyphysics@gmail.com, pal.sayan566@gmail.com, anb7101@gmail.com, biswajit@bose.res.in}, Sayan Kumar Pal$^*$, Aritra N. Bose$^*$, Biswajit Chakraborty$^*$}
\date{}

\maketitle

\begin{abstract}
We construct an effective commutative Schr\"odinger equation in Moyal space-time in $(1+1)$-dimension where both $t$ and $x$ are operator-valued and satisfy $\left[ \hat{t}, \hat{x} \right] = i \theta$. Beginning with a time-reparametrised form of an action we identify the actions of various space-time coordinates and their conjugate momenta on quantum states, represented by Hilbert-Schmidt operators. Since time is also regarded as a configuration space variable, we show how an `induced' inner product can be extracted, so that an appropriate quantum mechanical interpretation is obtained. We then discuss several other applications of the formalism developed so far.

\end{abstract}

\section{Introduction}

The nature of time in quantum mechanics remains quite intriguing till today. This fact is obvious to any practitioners of quantum theory, who must have observed the asymmetrical role played by space and time coordinates, in the sense that time is regarded as an $c$-number evolution parameter and not elevated to the level of operators, unlike the spatial coordinates. As far as we are aware this point was emphasised long back by none other than Pauli \cite{Pau}, who argued that if time `$t$' is also elevated to the level of operators then the energy spectrum will be continuous taking values in the entire interval $\left( -\infty, \infty \right)$ (we will recall this argument in the sequel). Apart from this, there is a huge literature on this area. See for example, \cite{Rom, Muga, Bru, But} and references therein. Another significant work was due to T. D. Lee who had considered time as a dynamical variable in \cite{Lee} for non-relativistic field theories and path integral over time has been formulated. From a different perspective, coherent state quantization of time function for a free particle has been introduced in \cite{Gaz}, \cite{Ah}, \cite{Gaz2}. Reparametrization of time was also introduced earlier in \cite{Un, Rob, RP}.

In a different context, a very strong plausibility argument was provided by Doplicher \emph{et. al.} \cite{Dop} that localization of an event down to Planck length scale $\left( l_p = \sqrt{\frac{\hbar G}{c^3}} \sim 10^{-33} cm \right)$ is virtually impossible if the basic tenets of general relativity and quantum theory persists to be valid even to that scale, as any process of localization will give rise to a gravitational collapse. One of the plausible ways to evade this kind of collapse is to impose noncommutative algebra between both space and time coordinates. One of the simplest such model is a Moyal space-time described by
\begin{equation} \label{intro1}
\left[ \hat{x}^\mu, \hat{x}^\nu \right] = i \theta^{\mu\nu}
\end{equation}
Models, where space-time coordinates satisfy similar or more general type of noncommutative algebra, where time is necessarily operator-valued were also considered. For example, scattering theory is formulated and an outline of quantum field theory has been provided in \cite{Bal2}. In \cite{Lu}, particle dynamics on Snyder spaces have been studied. We apologise for other major omissions in citation, if any. In this paper we intend to study $(1+1)$-dimensional noncommutative non-relativistic quantum mechanics where both time and space coordinates are operator-valued and satisfy the following commutation relation
\begin{equation} \label{intro2}
\left[ \hat{t}, \hat{x} \right] = i \theta ~~;~~ \theta > 0
\end{equation}
For $\theta < 0$, we can flip the sign of $\hat{x} \to -\hat{x}$ to restore \eqref{intro2}. However, note that this parity symmetry is not respected here unlike time reversal symmetry. We shall show subsequently, that by treating time as an operator here with the commutation relation \eqref{intro2} has no bearing with the above-mentioned Pauli's objection. However, in absence of any real parameter taking care of time evolution the problem becomes quite non-trivial even from a conceptual standpoint. On the other hand, in the technical level, there were claims that quantum field theories based on \eqref{intro1} are necessarily non-unitary \cite{Gom,Chai}. However, in a subsequent publication \cite{Dop2}, Doplicher and his collaborators have shown that it is quite possible to formulate quantum field theories which are ultraviolet finite to all orders. The point they emphasised was that the evolution parameter should not be identified with the eigenvalues of $\hat{t}$. Although they coalesce in the commutative limit, their conceptual distinction in the noncommutative case should be taken care of throughout the analysis. Indeed Balachandran \emph{et. al.}, following \cite{Dop} was able to formulate non-commutative quantum mechanics \cite{Bal} appropriate for \eqref{intro2} and study various applications.

Here we start with a time re-parametrised invariant form of a non-relativistic action and obtain the Schr\"odinger equation both in the commutative ($\theta = 0$ in \eqref{intro2}) and eventually to noncommutative quantum mechanics. This was primarily inspired by the earlier works by Deriglazov \cite{Deri}. We then try to construct an effective commutative Schr\"odinger equation by making use of the coherent state basis in a Hilbert-Schmidt operator formalism developed earlier in \cite{SCGV}. Afterwards we show how the appropriate inner product, necessary for conventional probability interpretation to go through, can be extracted from the Hilbert space of Hilbert-Schmidt operators. As applications of this formalism we investigate the noncommutative effect in (i) the time evolution of a Gaussian packet in momentum space, (ii) harmonic oscillator, (iii) Ehrenfest theorem along with various uncertainty relations and finally the deformation in Fermi's golden rule in presence of time dependent potential.

The paper is organised as follows. We begin with a discussion on our time re-parametrisation scheme in section \ref{sec2}. In section \ref{sec3}, we introduce Hilbert-Schmidt operator formalism for noncommutative quantum mechanics, the one we will adopt in this analysis. Then, we will briefly discuss the proposed techniques to extract an `effective' commutative theory from the noncommutative operatorial formalism in section \ref{sec4}. We then apply this formalism to several quantum systems in later sections. In section \ref{sec5} we discuss the time evolution of a free particle Gaussian wave packet in momentum space and in section \ref{sec6} the harmonic oscillator problem has been analysed. We then proceed on to study the possible manifestations of space-time noncommutativity in different quantum systems. In section \ref{sec7} the status of expectation values of different operators, their uncertainty relations and modifications in Ehrenfest theorem has been analysed. Section \ref{sec8} deals with the modification in the transition probability in presence of time-dependent potentials. In section \ref{sec9}, we discuss the possible modifications in Galilean algebra and Galilean generators in Moyal space time. Finally, we conclude in section \ref{sec10}.

\section{Quantum Mechanics in (1+1) dimension} \label{sec2}

In this section, we begin with a brief review of time-reparametrised invariant form of the action \cite{Tet}, where the time is treated as a configuration space variable in addition to the position. For this, we essentially follow Deriglazov \emph{et al} \cite{Deri}. We begin by considering the action of the non-relativistic particle in the presence of the potential $V(x,t)$ (can depend on time t also) in one dimension as
\begin{equation}
S[x(t)]=\int_{t_1}^{t_2} dt L\left(x,\frac{dx}{dt}\right),~~~L=\frac{1}{2}m\left(\frac{dx}{dt}\right)^2 - V(x,t).
\end{equation}
Now taking the new parameter $\tau$ as an evolution parameter we parametrise the time as $t=t(\tau)$ along with the position variable $x=x(\tau)$ and treat both as configuration space variables. We just require $t(\tau)$ to be a monotonically increasing function of $\tau$. With this the above action can be re-written as
\begin{equation}
S[x(\tau),t(\tau)]=\int_{\tau_1}^{\tau_2} d\tau L_\tau(x,\dot{x},t,\dot{t}),~~~L_\tau(x,\dot{x},t,\dot{t})=\frac{1}{2}m\frac{\dot{x}^2}{\dot{t}} -\dot{t}V(x,t),
\end{equation}
where the over-head dot now indicates differentiation w.r.t. $\tau$ i.e. $\dot{t}=\frac{dt}{d\tau}$, $\dot{x}=\frac{dx}{d\tau}$. Clearly the canonical momenta corresponding to the configuration space variables $t(\tau)$ and $x(\tau)$ are now given by
\begin{equation}
p_x =\frac{dL_\tau}{d\dot{x}}=m\frac{\dot{x}}{\dot{t}}=m\left(\frac{dx}{dt}\right)
\end{equation}
and
\begin{equation} \label{n1}
\begin{split}
p_t & = \frac{dL_\tau}{d\dot{t}}=-\frac{1}{2} m\frac{\dot{x}^2}{\dot{t}^2}-V(x,t)=-\frac{1}{2} m\left(\frac{dx}{dt}\right)^2-V(x,t) \\
 & =-\frac{p_x^2}{2m}-V(x,t)=-H \
\end{split}
\end{equation}
with $H = \frac{p_x^2}{2m} + V(x,t)$.\\

This indicates the presence of a primary constraint given by
\begin{equation}
\phi=p_t + H\approx 0.\label{5}
\end{equation}
Here $\approx $ refers to the equality in the weak sense \cite{Dirac, Tet}. The Legendre transformed Hamiltonian $H_\tau$ corresponding to $L_\tau$ becomes proportional to this constraint and so also vanishes weakly:
\begin{equation}
H_\tau = p_t\dot{t}+p_x\dot{x}-L_\tau=\dot{t}\phi \approx 0.\label{nn}
\end{equation} 
Clearly, there are no secondary constraints and this being the only constraint, it is first class in Dirac`s classification of constraints and therefore generates gauge transformation. This implies that $\tau$-evolution now can be identified with unfolding of this gauge transformation. The corresponding quantum theory is now constructed by elevating all the phase space variables ($t,x,p_t,p_x$) to the level of operators satisfying Heisenberg algebra (in the unit $\hbar = 1$):
\begin{equation}
[\hat{t},\hat{x}]=0=[\hat{p_t},\hat{p_x}],~~[\hat{t},\hat{p_t}]=i=[\hat{x},\hat{p_x}].\label{A2}
\end{equation}
We then look for a Hilbert space, furnishing a representation of this algebra. This is clearly $L^2(\mathcal{R}^2)$ as the configuration space is now two dimensional. We now introduce the spatio-temporal simultaneous eigen basis $\ket{x,t}$ of the commutating $\hat{t}$ and $\hat{x}$ operators satisfying
\begin{equation}
\hat{t}\ket{x,t}=t\ket{x,t},~~~\hat{x}\ket{x,t}=x\ket{x,t},\label{A3}
\end{equation}
along with orthonormality and completeness relation as,
\begin{equation}
\left\langle x^\prime, t^\prime | x,t \right\rangle = \delta (x^\prime - x)\delta(t^\prime -t)~;~\int dtdx \ket{x,t}\bra{x,t}=1.
\end{equation}
The representations of phase space operators are given as,
\begin{equation} \label{n2}
\begin{split}
\left\langle x,t | \hat{x} | \psi \right\rangle = x\left\langle x,t |\psi \right\rangle , & \left\langle x,t |\hat{t} |\psi \right\rangle = t \left\langle x,t |\psi \right\rangle\\
\left\langle x,t |\hat{p}_x |\psi \right\rangle = -i\partial_x \left\langle x,t |\psi \right\rangle , & \left\langle x,t |\hat{p}_t |\psi \right\rangle = -i\partial_t \left\langle x,t |\psi \right\rangle \
\end{split}
\end{equation}
where $\psi (x,t)=\left\langle t,x|\psi\right\rangle \in L^2(\mathcal{R}^2)$ and can be formally identified with the wave function. The corresponding norm is now given by
\begin{equation}
\left\langle\psi |\psi\right\rangle =\int dtdx~ \psi^\ast(x,t)\psi(x,t) < \infty .\label{nn2}
\end{equation}
Finally the time-dependent Schr\"odinger equation is obtained by demanding that physical states i.e. these $\ket{\psi}_{phy}$'s be gauge invariant. In other words, the first class constraint annihilates the physical state of the system:
\begin{equation}
\hat{\phi}\ket{ \psi}_{phy}=(\hat{p_t} + \hat{H})\ket{ \psi}_{phy}=0.\label{nn1}
\end{equation}
This readily yield time dependent Schr\"odinger equation by taking overlap with $\ket{x,t}$ and using (\ref{n1},\ref{n2}) as
\begin{equation}
i\frac{\partial}{\partial t} \psi(x,t)=\left( -\frac{1}{2m} \frac{\partial^2}{\partial x^2} + V(x,t)\right) \psi (x,t).\label{A7}
\end{equation}
Note that it is independent of the evolution parameter $\tau$, as its $\tau$-evolution is frozen, as can be easily seen by using (\ref{nn},\ref{nn1}). The usual probabilistic interpretation in quantum mechanics is then recovered by replacing the inner-product 
\begin{equation}
\left\langle\psi |\phi\right\rangle \equiv\int dtdx~ \psi^\ast(x,t)\phi(x,t),\label{in2}
\end{equation}
appropriate for the norm (\ref{nn2}) for the Hilbert space $L^2(\mathcal{R}^2)$ to that of $L^2(\mathcal{R}^1)$ i.e. by the one, which involves only a spatial integration at a constant time slice:
\begin{equation}
\left\langle\psi |\phi\right\rangle_t : =\int_{t} dx ~\psi^\ast(x,t)\phi(x,t).\label{in1}
\end{equation}

We shall refer to this as ``induced inner product". Clearly, normalizable states with $L^2(\mathcal{R}^1)$ inner product (\ref{in1}) may not be so w.r.t. that of $L^2(\mathcal{R}^2)$ (\ref{in2}): $L^2(\mathcal{R}^2)\subset L^2(\mathcal{R}^1)$. As an example, we may consider the typical example of a stationary state like $\psi (x,t) = e^{-iEt} \phi (x)$. Finally, note that the self-adjoint-ness of the derivative representation of $\hat{p}_t=-i\partial_t$ in (\ref{n2}) is no longer valid in the Hilbert space $L^2(\mathcal{R}^1)$ with associated inner product \eqref{in1}, as it is not sensible to demand that $\abs{\psi(x,t)} \to 0$ as $\abs{t} \to \infty$. In contrast, in $L^2(\mathcal{R}^2)$, this would have allowed one to carry out integration by parts and drop boundary terms. Indeed, this is closely related to the original Pauli's objection \cite{Pau} in regard to the elevation of $\left( \hat{t}, \hat{p}_t \right)$ to the level of operators. His arguments were quite simple, which we recall here very briefly. Considering an energy eigenstate $\ket{E}$ satisfying $\hat{p}_t \ket{E} = - \hat{H} \ket{E} = - E \ket{E}$ \eqref{nn1}, the state $e^{i \alpha \hat{t}} \ket{E}$ too will be an eigenstate $\ket{E - \alpha}$ with energy eigenvalue $(E - \alpha)$, where $\alpha$ is an arbitrary real parameter, allowing the spectrum of the Hamiltonian $H$ to be continuum with values in the entire range $\left( -\infty, \infty \right)$. Particularly, this is in direct conflict with the existence of systems where energy is bounded from below. Although, there were some attempts to to evade this hurdle \cite{Olkh}, we are not going to pursue this approach and rather follow the conventional approach, where $\hat{p}_t$ is now excluded from the phase space variables, along with $\hat{t}$. The latter, when `demoted' to a $c$-number parameter, is now identified with the new evolution parameter with $\left( -i\partial_t \right)$ having no association with $\hat{p}_t$ anymore, so that (\ref{A7}) has now the status of a postulate.

\section{Quantum mechanics with space-time noncommutativity} \label{sec3}
We now provide a formulation of non-relativistic quantum mechanics in Moyal space-time, where the Heisenberg algebra (\ref{A2}) is replaced by the so-called non-commutative Heisenberg algebra (NCHA):
\begin{equation}
[\hat{t},\hat{x}]=i\theta,~~[\hat{p}_t,\hat{p}_x]=0;~~[\hat{t},\hat{p_t}]=i=[\hat{x},\hat{p_x}],\label{A10}
\end{equation}
with $\theta$ being the noncommutative parameter. Our formulation is in some sense, an extension of the Hilbert-Schmidt operatorial formulation of quantum mechanics \cite{SCGV, Roh}, where time was the usual evolution parameter and noncommutative algebra between the operator-valued position coordinate variables for 2D Moyal plane was only considered. In the spirit of the previous section, here too we consider time as a configuration space variable in the beginning, so that in analogy with Moyal plane, we introduce an auxiliary Hilbert space
\begin{equation}
\mathcal{H}_c = Span\left\lbrace |n\rangle = \frac{(b^\dagger)^n}{\sqrt{n!}} |0\rangle ~;~ b = \frac{\hat{t} + i\hat{x}}{\sqrt{2\theta}} \right\rbrace,\label{b1}
\end{equation}
furnishing a representation of just the coordinate algebra: $[\hat{t},\hat{x}]=i\theta$ or, equivalently that of $[\hat{b}, \hat{b}^\dagger]=1$. A unitary representation of the the entire NCHA (\ref{A10}) is then furnished by the Hilbert space $\mathcal{H}_q$ of Hilbert Schmidt operators,
\begin{equation} \label{def_H_q}
\mathcal{H}_q = \left\lbrace \psi(\hat{t},\hat{x})\equiv |\psi(\hat{t},\hat{x})) ;~ tr_c\left(\psi^\dagger (\hat{t},\hat{x})\psi(\hat{t},\hat{x}) \right) < \infty \right\rbrace
\end{equation}
acting on $\mathcal{H}_c$. They are essentially the elements of the algebra of polynomials generated by $(\hat{t},\hat{x})$ and correspond to compact and trace-class operators. The appropriate actions for the phase space operators are given by:
\begin{equation}
\hat{T}  \psi(\hat{t},\hat{x})  = \hat{t} \psi(\hat{t},\hat{x}),~~\hat{X}  \psi(\hat{t},\hat{x})  =\hat{x} \psi(\hat{t},\hat{x}) ,\label{19}
\end{equation}
\begin{equation}
 \hat{P_x}  \psi(\hat{t},\hat{x})  = - (\theta^{-1})  [\hat{t}, \psi(\hat{t},\hat{x})] ,~~\hat{P_t}  \psi(\hat{t},\hat{x})  =  (\theta^{-1})  [\hat{x}, \psi(\hat{t},\hat{x})].\label{20}
\end{equation}
Here the capital letters $\hat{T}$ and $\hat{X}$ have been used in place of $\hat{t}$ and $\hat{x}$ respectively to distinguish their domains of action viz, $\mathcal{H}_q$ and $\mathcal{H}_c$ respectively; the former pairs can be regarded as representations of the latter. It can be easily checked that all these phase space operators are self adjoint with respect to the inner product
\begin{equation}
(\psi |\phi)=tr_{\mathcal{H}_c}(\psi^\dagger \phi)~~~~~ \forall~ \psi, \phi \in \mathcal{H}_q. \label{A14}
\end{equation}
It is now clear that in view of $\theta \neq 0$, that a counter part of the common eigenstate $\ket{x,t}$ (\ref{A3}) can not exist. However, since $\hat{P}_x$ and $\hat{P}_t$ still commutes, common eigenstate $|p,E)$ in the fourier space of these operator satisfying
\begin{equation}
\hat{P}_x|p,E)=p|p,E),~~\hat{P}_t|p,E)=-E|p,E)\label{A15}
\end{equation}
should exist. Indeed, it can be easily checked that the following state 
\begin{equation}
|p,E)=\sqrt{\frac{\theta}{2\pi}}e^{-i(E\hat{t}-p\hat{x})}\label{23}
\end{equation}
satisfies (\ref{A15}), apart from orthonormality and completeness relation
\begin{equation} \label{24}
(p,E|p^\prime ,E^\prime)=\delta (p-p^\prime) \delta (E-E^\prime);~~~\int dp dE~ |p,E)(p,E|=1_q.
\end{equation}

It should be noted at this stage that the inner product (\ref{A14}) corresponds to (\ref{in2}) for the commutative ($\theta = 0$) case. Further, note that the vectors in $\mathcal{H}_q$ are being denoted by round kets $|.)$, in contrast to angular $\ket{.}$ in $\mathcal{H}_c$. The stage is ready to write down Schr\"odinger equation-the counter part of the (\ref{A7}). For that, we start with(\ref{5}), assuming just that this equation still holds, even in the presence of noncommutativity  
\begin{equation} \label{eq25}
(\hat{P}_t + \hat{H}) \hat{\psi}(\hat{x},\hat{t}) =0,~~~~~~~\hat{H}=\frac{\hat{P}_x^2}{2m} + V(\hat{X},\hat{T}).
\end{equation}

We further assume that $V(\hat{X}, \hat{T})$ is hermitian with suitable operator-ordering. Now using (\ref{19},\ref{20}), we can write down the abstract operator form of the Schr\"odinger equation
\begin{equation}
\frac{1}{2m\theta}[\hat{t},[\hat{t},\hat{\psi}]] + [\hat{x},\hat{\psi}] +V(\hat{x},\hat{t})\hat{\psi} =0.\label{sc}
\end{equation}

\paragraph*{Continuity equation\\}

Proceeding as in the commutative case, we multiply both sides of (\ref{sc}) by $\psi^\dagger$  to get
\begin{equation}
\frac{1}{2m\theta}\hat{\psi}^\dagger[\hat{t},[\hat{t},\hat{\psi}]] + \hat{\psi}^\dagger[\hat{x},\hat{\psi}] + \hat{\psi}^\dagger V(\hat{x},\hat{t})\hat{\psi} =0.
\end{equation}
The hermitian conjugate of the above equation is given by
\begin{equation}
\frac{1}{2m\theta}[\hat{t},[\hat{t},\hat{\psi}^\dagger]]\hat{\psi} - [\hat{x},\hat{\psi}^\dagger]\hat{\psi} + \hat{\psi}^\dagger V(\hat{x},\hat{t})\hat{\psi} =0.
\end{equation}
Now it is quite straight forward to see that the difference between these pair of equation yields the abstract operator form of the continuity equation as,
\begin{equation}
[\hat{x},\rho] + [\hat{t},J]=0.\label{29}
\end{equation}
where
\begin{equation}
\rho =\hat{\psi}^\dagger \hat{\psi},~~~~J=\frac{1}{2m\theta}\left(\hat{\psi}^\dagger [\hat{t},\hat{\psi}]- [\hat{t},\hat{\psi}^\dagger]\hat{\psi}\right)\label{b5}
\end{equation}
should now correspond to the probability density $\rho$ and probability current $J$ respectively, if an appropriate inner product, i.e. the counter part of (\ref{in1}) can be introduced in the presence of noncommutativity also. We take up this case in the next section.

\section{Recovery of effective commutative theory} \label{sec4}

In this section, we would like to construct an explicit space-time coordinate representation of the above operatorial version of Schr\"odinger equation (\ref{sc}) and continuity equation (\ref{29}). As mentioned earlier, the non-existence of common eigenstate of $\hat{T}$ and $\hat{X}$ operators, the counter part of (\ref{A3}) in the view of their noncommutativity, makes the task a bit non-trivial. Particularly, we need the analogue of the inner product (\ref{in1}), involving only the spatial integration on a fixed time slice to formulate an effective and equivalent commutative quantum theory. Clearly, the best choice is to use the coherent state
\begin{equation}
\ket {z} = e^{-\bar{z}b+zb^\dagger}\ket {0}=e^{-\frac{1}{2}|z|^2}e^{zb^\dagger}\ket{0} \in \mathcal{H}_c ,\label{b3}
\end{equation}
which is an eigenstate of the annihilation operator $b$ (\ref{b1})
\begin{equation}
b\ket{z}=z\ket{z}\label{b2}
\end{equation}
and is a maximally localised state in $\mathcal{H}_c$: $\bigtriangleup t \bigtriangleup x =\frac{\theta}{2}$. Here $z$ is an arbitrary dimension-less complex number and can be splitted into real and imaginary parts as,
\begin{equation}
z=\frac{t+ix}{\sqrt{2\theta}}\label{b4}
\end{equation}
as suggested by (\ref{b1},\ref{b2}) so that $t$ and $x$ can be regarded as effective commutative time and space coordinates. The above coherent state $\ket{z}$ \eqref{b3} can then also be labelled, alternatively, by this pair of parameters $\ket{z}=\ket{x,t}$. We now construct a basis in $\mathcal{H}_q$ by taking the outer product of $\ket{z}$ (\ref{b3}):
\begin{equation} \label{vbasis}
|z,\bar{z})\equiv |z)=\ket{z}\bra{z} = \sqrt{2\pi\theta} ~ |x,t)_V \in \mathcal{H}_q ~~;~~B|z,\bar{z})=z|z,\bar{z}),
\end{equation}
where the annihilation operator $B=\frac{\hat{T}+i\hat{X}}{\sqrt{2\theta}}$ can be regarded as the representation of the operator $b$ on $\mathcal{H}_q$. The use of the subscript $V$ and the pre-factor $\sqrt{2\pi\theta}$ will be justified soon. Then the space-time representation, in coherent state basis, referred to as ``symbols" of the abstract state $|\psi(\hat{x},\hat{t}))\in \mathcal{H}_q$ \eqref{def_H_q}, is then obtained by taking the overlap with $|z)$ to get, using (\ref{A14}),
\begin{equation} \label{eq35}
\psi(x,t) \equiv ~_V(x,t|\psi) = \frac{1}{\sqrt{2\pi\theta}} (z|\psi(\hat{x},\hat{t}))= \frac{1}{\sqrt{2\pi\theta}} \left\langle z|\psi(\hat{x},\hat{t})|z\right\rangle
\end{equation}
In particular, choosing $\psi(\hat{x},\hat{t})=\psi(b,b^\dagger)= \sqrt{2\pi\theta} b$, the corresponding wave function is (\ref{b4}) itself.\footnote{Although, by itself it is not a compact operator and therefore not a Hilbert-Schmidt operator, it can, however, be regarded as an element of the multiplier algebra.} Now to obtain a similar representation for $\rho =\hat{\psi}^\dagger(\hat{x},\hat{t})\hat{\psi}(\hat{x},\hat{t})$ (\ref{b5}) we need to discuss the corresponding representation of a generic composite operator. For this purpose, first note that $\mathcal{H}_q$ has the structure of algebra: the operator product of any pair of operators $\hat{\psi}(\hat{x},\hat{t})$ and $\hat{\phi}(\hat{x},\hat{t})$ of $\mathcal{H}_q$ is an another element of $\mathcal{H}_q$; it is closed under the multiplication map $\mu$:

\begin{eqnarray}
\mu &:& \mathcal{H}_q \otimes \mathcal{H}_q \rightarrow \mathcal{H}_q\nonumber\\
 & & \mu \left( \psi(\hat{x},\hat{t}) \otimes \phi(\hat{x},\hat{t}) \right) = \psi(\hat{x},\hat{t})\phi(\hat{x},\hat{t}).
\end{eqnarray}
Further, it is shown in \cite{Basu} that the representation of composite operators is identical to the one obtained by composing the representation of individual operators through Voros star product. In other words, the operator algebra is isomorphic to symbol algebra iff the elements of the latter is composed through Voros star product:
\begin{equation}
\left( z|\psi(\hat{x},\hat{t}) \, \phi(\hat{x},\hat{t})\right)= (z|\psi(\hat{x},\hat{t})) \star_V (z|\phi(\hat{x},\hat{t}))\label{b7}
\end{equation}
where the Voros star product $\star_V$ can be written by making use of (\ref{b4}), as
\begin{equation} \label{b8}
\star_V =e^{\overleftarrow{\partial_z}\overrightarrow{\partial_{\bar{z}}}}= e^{\frac{i}{2}\theta(-i \delta_{ij} + \epsilon_{ij} ) \overleftarrow{\partial_i} \overrightarrow{\partial_j} } = e^{\frac{\theta}{2} \overleftarrow{\partial_i} \overrightarrow{\partial_i} } \, \star_M ; ~ \star_M = e^{\frac{i}{2}\theta \epsilon_{ij} \overleftarrow{\partial_i} \overrightarrow{\partial_j} } ; ~ i,j=0,1 ; ~ x^0=t, ~x^1=x ; ~\epsilon_{01}=1.
\end{equation}

Here we have also displayed how the Moyal star product $\star_M$ is related to $\star_V$. Returning to the expansion of $\rho$ \eqref{b5}, in particular, we see that this yields, on using \eqref{eq35},
\begin{equation} \label{b6}
\begin{split}
\rho(x,t) & \equiv ~_V(x,t|\rho(\hat{x},\hat{t})) = \sqrt{2\pi\theta} ~_V(x,t|\hat{\psi}^\dagger(\hat{x},\hat{t})) \star_V \, _V(x,t|\hat{\psi}(\hat{x},\hat{t})) = \sqrt{2\pi\theta} ~ \psi^\ast(x,t) \star_V \psi(x,t).
\end{split}
\end{equation}
Further note that it is only for Voros star product that the positive definiteness property of $\rho(x,t)$ can be ensured. This can be easily seen by using \eqref{b7}, where we can write $\rho(x,t)$ in a manifestly positive definite form
\begin{equation} \label{eq40}
\rho(x,t) = \frac{1}{\sqrt{2\pi\theta}} \psi^\ast(z,\bar{z})\star_V \psi(z,\bar{z}) = \frac{1}{\sqrt{2\pi\theta}} \sum_{n=0}^\infty \frac{1}{n!}|\partial_z^n \psi(z,\bar{z})|^2 \geqslant 0.
\end{equation}
This is in contrast with Moyal star product $\star_M $ \eqref{b8} and thus makes it essential to use Voros star product to allow us to have the probability interpretation to go through. This is reminiscent of quantum mechanics in 2D Moyal plane \cite{Basu}, where the Voros basis was compatible with POVM, rather than the Moyal basis. Note that, here we are referring to the coherent state basis $|z) = \sqrt{2\pi\theta} |x,t)_V$ \eqref{vbasis} as the Voros basis, as this is associated with Voros star product. A similar basis associated to Moyal star product, the so-called Moyal basis was also constructed in \cite{Basu} and can easily be carried out here as well, but we won't need it, as positive definitiveness of $\rho(x,t)$ can not be ensured here, as mentioned above. The rest of the paper therefore, deals with only Voros star product and its associated basis. Henceforth we shall thus omit the subscript $V$.

It is now quite straight forward to see that the resolution of identity takes the following form:
\begin{equation} \label{idn}
\int \frac{d^2z}{\pi}~ |z) \star (z| = \int dtdx \, |x,t) \star (x,t| = 1_q,
\end{equation} 
as can be proved easily by sandwiching it in the orthonormality relation (\ref{24}) of energy momentum eigenstate $|p,E)$ (\ref{23}) and making use of the overlap
\begin{equation} \label{eq42}
\frac{1}{\sqrt{2\pi\theta}} (z|p,E) = (x,t|p,E) = \frac{1}{2\pi} e^{-\frac{\theta}{4}(E^2 + p^2)} e^{-i(Et-px)}.
\end{equation}
This suggests that the inner product between any pair of elements in the Hilbert space of symbols corresponding to the elements of $\mathcal{H}_q$ should also be defined through Voros star product :
\begin{equation} \label{innpro_voros}
(\psi|\phi) = \int dtdx ~ \psi^\ast(x,t) \star \phi(x,t)
\end{equation}
This is the counterpart of \eqref{in2} in ``commutative" quantum mechanics to which it reduces to in the limit $\theta \to 0$. Also note that the overlap of the basis $|x,t)$ \eqref{vbasis} and its counterpart is given by
\begin{equation} \label{a3}
(x^\prime, t^\prime | x, t ) = \frac{(z^\prime|z)}{2\pi\theta} = \frac{\abs{\bra{z^\prime}\ket{z}}}{2\pi\theta} = \delta_{\sqrt{\theta}} (t^\prime - t) \delta_{\sqrt{\theta}} (x^\prime - x)
\end{equation}
where
\begin{equation} \label{a4}
\delta_\sigma (x) = \frac{1}{\sigma\sqrt{2\pi}} e^{-\frac{x^2}{2\sigma^2}} ~;~~ \int dx \, \delta_\sigma (x) = 1
\end{equation}
Finally note that the star product intertwines $t$ and $x$ dependence and $\left( \delta_{\sqrt{\theta}} (t^\prime - t) \delta_{\sqrt{\theta}} (x^\prime - x) \right)$ - as a whole, plays the role of Dirac's $\delta$-distribution in our noncommutative  space-time, provided they are composed with the star product. This can be seen quite transparently from the derivation of the following identity, by making use of \eqref{eq42}
\begin{equation} \label{5a1}
\int dt^\prime dx^\prime \, \left( \delta_{\sqrt{\theta}} (t - t^\prime) \, \delta_{\sqrt{\theta}} (x - x^\prime) \right) \star^\prime \psi \left( x^\prime, t^\prime \right) = \psi \left( x, t \right) ~~;~ \psi(x,t) = (x,t|\psi) = \int dE dp \, (x,t|E,p)(E,p|\psi)
\end{equation}
where it is essential to retain both $\delta_{\sqrt{\theta}} (t)$ and $\delta_{\sqrt{\theta}} (x)$ together. This, in turn, can be seen easily by making use of the identity :
\begin{equation} \label{7b1}
\int dt' dx' \, \delta_{\sqrt{\theta}} (t - t^\prime) \, \delta_{\sqrt{\theta}} (x - x^\prime) \star^\prime e^{-i \left( Et^\prime - px^\prime \right)} = e^{-i \left( Et - px \right)}
\end{equation}
where $\star^\prime$ indicates that the relevant derivative involve $t^\prime$ and $x^\prime$. Besides, to recover effective commutative theory with usual interpretations of quantum mechanics we recall Pauli's objection and exclude `$t$' and `$p_t$' from the phase space variables. This, however, \emph{does not} imply that $\hat{t}$ is no longer an operator; it still satisfies $[\hat{t}, \hat{x}] = i \theta$ \eqref{A10}, but the pair of commutators involving $\hat{P}_t$ in (\ref{A10}, \ref{20}) are disregarded. Particularly, the operator $\frac{1}{\theta} \, ad\,\hat{x}$ is no longer identified with $\hat{P}_t$. On the other hand $e^{i\alpha\hat{t}}$ generates translation in $\mathcal{H}_c$ in the sense that its action on an eigenstate $\ket{a}$ of $\hat{x}$, satisfying $\hat{x}\ket{a} = a\ket{a}$ yields a shifted eigenstate of $\hat{x}$ since $\hat{x} \, \left( e^{i\alpha\hat{t}} \ket{a} \right) = \left( a + \alpha \theta \right) \left( e^{i\alpha\hat{t}} \ket{a} \right)$, so that we can write $e^{i\alpha\hat{t}} \ket{a} = \ket{a + \alpha\theta}$. This in turn implies, on taking outer product, that $\dyad{a + \alpha \theta}{a + \alpha \theta} = e^{i\alpha\hat{t}} \dyad{a}{a} e^{-i\alpha\hat{t}}$. In its infinitesimal version, this indeed enables us to identify $\hat{P}_x$ with $\left(-\frac{1}{\theta}\right) ad\,\hat{t}$, as it occurs in \eqref{20}. Therefore, finally again the Schr\"odinger equation \eqref{sc} has the status of a postulate. Note that we can regard the basis $|x,t)$ as ``quasi-orthonormal bases", as Gaussian function \eqref{a4} can be regarded as some sort of ``regularised Dirac's $\delta$-distribution", in the sense that $\delta_\sigma (x) \to \delta (x)$ as $\sigma \to 0$. It is quite transparent at this stage that all these expressions of the previous section i.e. their commutative counterparts are reproduced in the limit $\theta \to 0$.

We now need to extract the conventional quantum mechanical inner product from (\ref{A14}, \ref{innpro_voros}) i.e. the analogue of \eqref{in1} from \eqref{in2}. To that end, let us make use of the completeness relation \eqref{24} satisfied by the basis $|p,E)$, and introduce a projection operator $\mathcal{P}_E$, at constant energy surface $E$ as,
\begin{equation} \label{cal_p_E}
\mathcal{P}_E = \int dp \, |p,E)(p,E| ~~;~~ \mathcal{P}_{E^\prime} \mathcal{P}_E = \mathcal{P}_E \delta(E^\prime - E)
\end{equation}

Using this projection operator $\mathcal{P}_E$ we can introduce the projected state $|\psi)_E = \mathcal{P}_E |\psi)$ and its coherent state representation :
\begin{equation} \label{eq50}
\psi_E (x,t) \equiv (x,t|\psi)_E = \int dp \, (x,t|p,E)(p,E|\psi) = \frac{1}{\sqrt{2\pi}} \int dp ~ e^{-i \left( Et - px \right)} \, e^{-\frac{\theta}{4} \left( E^2 + p^2 \right)} \, \psi_E (p)
\end{equation}
where we define $\psi_E(p) \equiv \frac{1}{\sqrt{2\pi}} ( p,E | \psi )$. If this is regarded as a stationary state\footnote{Typically, the stationary states will correspond to a discrete set of energy levels of a bound system, where the integration over $E$ i.e. $\int dE \cdots$ is to be replaced by summation $\sum\limits_E$, so that we can write for a general state $|\psi)$,
\begin{equation} \label{eqb3}
\psi (x,t) = \sum\limits_n C_n \psi_\theta (x,t)
\end{equation}
These coefficients $C_n$ now have a dimension $[L^{-1}]$. Correspondingly, we need to replace $\delta(E^\prime - E)$ by $\delta_{E^\prime E}$ in \eqref{cal_p_E}.}, then the time evolution is of the form of commutative quantum mechanics with the associated parameter $t$ being just a $c$-number, as $t$-dependence factors out in the usual manner. But one should keep in mind that this $t = \sqrt{\frac{\theta}{2}} \bra{z} \hat{b} + \hat{b}^\dagger \ket{z}$ and therefore is an expectation value and more precisely, the time evolution parameter $\tau$ (not the same one, that appeared previously) is given by the unitary operator $U(\tau) = e^{-iH\tau}$ for a time independent $H$. In other words, the evolution of the basis $|x,t)$ \eqref{vbasis} is now given by $|x, t+\tau ) = e^{i\hat{H}\tau} |x,\tau)$. But the point that needs to be emphasised is that here $\tau$ should not be identified as coordinate time $t$, \cite{Dop, Bal} i.e. as an eigenvalue of $\hat{t}$. In fact, $\tau$ by itself is not subjected to any quantum fluctuations and the coherent state $|x,t)$ can be regarded as an analogue of position basis in Heisenberg picture. States like $|\psi)_E$ \eqref{eqb3} span a subspace $\mathcal{H}_q(E) : \mathcal{H}_q(E) \subset \mathcal{H}_q$, with energy $E$ and subspaces associated  with distinct different energies are orthogonal to each other, thus splitting $\mathcal{H}_q$ into an one-parameter family of non-overlapping subspaces, parametrised by energy $E$. Schematically we may therefore write $\mathcal{H}_q = \bigoplus\limits_{E} \mathcal{H}_q(E)$.

We now consider the inner product \eqref{eq42} for a pair of states $|\psi)_E$ and $|\phi)_E \in \mathcal{H}_q(E)$ by making use of \eqref{eq50} to get
\begin{equation} \label{5a2}
\int dt dx \, \psi^\ast_E (x,t) \star \phi_E (x,t) = \frac{e^{-\frac{\theta}{2}E^2}}{2\pi} \int dt dx dp dp' \, e^{-\frac{\theta}{4}\left(p^2 + p'^2\right)} \, \psi^\ast_E (p) \phi_E (p') \left( e^{i\left(Et - px\right)} \star e^{-i\left(Et - p'x\right)} \right)
\end{equation}

A straightforward computation shows that `$t$'-dependence cancels out even in the presence of star product, yielding a divergent integral which can be written in another equivalent form, where the $x$-integration is replaced by $p$-integration :
\begin{equation} \label{5a3}
\int dt dx \, \psi^\ast_E (x,t) \star \phi_E (x,t) = \int dt dp \, \psi^\ast_E (p) \, \phi_E (p)
\end{equation}
However, since $t$ should no longer be counted as a phase space variable, we introduce an ``induced" inner product by excluding $t$-integration and the finite integral over $x$ or $p$ :
\begin{equation} \label{5a4}
\left( \psi_E | \phi_E \right)_t := \int_t dx \, \psi^\ast_E (x,t) \star \phi_E (x,t) = \int_t dp \, \psi^\ast_E (p) \, \phi_E (p)
\end{equation}
where the presence of `$t$' at the bottom of the integral sign indicates that the integration has to be performed over a constant $t$-surface. The aforementioned orthogonality $(\psi_E | \phi_{E^\prime} ) = 0$ between pair of states belonging to different energy surfaces $E^\prime \neq E$ (\ref{24}, \ref{cal_p_E}) can now be established again through the general form of induced inner product
\begin{equation} \label{5a5}
\left( \psi, \phi \right)_t = \int_t dx \, \psi^\ast (x,t) \star \phi (x,t) ~~~~\forall~ |\psi), |\phi) \in \mathcal{H}_q
\end{equation}
by invoking the self-adjointness of the Hamiltonian operator $H$, as in conventional commutative quantum mechanics :
\begin{equation}
\left( \psi, H \phi \right)_t = \left( H \psi, \phi \right)_t
\end{equation}

We finally note that \eqref{5a4}, upon normalization, can be regarded as the non-commutative version of Parseval's theorem, when $| \phi_E) = |\psi_E)$. Further observe that one can introduce
\begin{equation} \label{pi_t}
\pi_t = \int_t dx ~| x,t ) \star ( x,t |, 
\end{equation}
the counterpart of $\mathcal{P}_E$ \eqref{cal_p_E}to re-write the inner product \eqref{5a5} as
\begin{equation} \label{9a2}
\left( \psi, \phi \right)_t = \left( \psi | \pi_t | \phi \right)
\end{equation}
which, however, satisfies only approximately a deformed version of projection operator identity for small $\theta$ :
\begin{equation} \label{7a2}
\pi_{t^\prime} \pi_t \approx \pi_{t^\prime} \, \delta_{\sqrt{\theta}} (t^\prime - t)
\end{equation}
This indicates that any pair of projection operators $\pi_t$ and $\pi_{t+\delta t}$ separated by a time interval $\delta t$ will not be orthogonal exactly.

We now try to find a coordinate representation of various phase space operators (\ref{19},\ref{20}) in this coherent state basis. For that let us first consider $\hat{T}$ and $\hat{X}$ operators, whose actions has been defined  in (\ref{19}) through the left action: $\hat{T}_L\psi = \hat{t}\psi$ and $\hat{X}_L\psi = \hat{x}\psi$. This is of course a matter of convention and we could have defined the right action as well:
\begin{equation}
\hat{T}_R\psi(\hat{x},\hat{t}) = \psi(\hat{x},\hat{t}) \, \hat{t};~\hat{X}_R\psi(\hat{x},\hat{t}) = \psi(\hat{x},\hat{t}) \, \hat{x}.
\end{equation}
Let us first determine the coordinate representation $\hat{X}_L$ in (\ref{19}) as an example. To that end consider $(x,t|\hat{X}_L \, \psi(\hat{x},\hat{t}))$, which on using (\ref{b7}) can be written as,
\begin{equation}
( x,t|\hat{X}_L \, \psi(\hat{x},\hat{t})) = \sqrt{2\pi\theta} \, ( x,t|\hat{x}) \star (x,t|\psi(\hat{x},\hat{t})) = \frac{1}{\sqrt{2\pi\theta}} \, \left\langle z|\hat{x}|z\right\rangle \star (z|\psi(\hat{x},\hat{t}))
\end{equation}
Finally making use of \eqref{b8} this readily yields
\begin{equation}
( x,t|\hat{X}_L \, \psi(\hat{x},\hat{t})) = X_\theta \, (x,t|\psi(\hat{x},\hat{t})) \equiv X_\theta \, \psi(x,t)
\end{equation}
where
\begin{equation}
X_\theta^L \equiv X_\theta =  \left[x+\frac{\theta}{2}(\partial_x-i\partial_t)\right].\label{47}
\end{equation}
Proceeding exactly in the same manner, we get
\begin{equation}
T_\theta^L \equiv T_\theta =  \left[t+\frac{\theta}{2}(\partial_t+i\partial_x)\right].\label{48}
\end{equation}
Taking a pair of $|\psi_1)$, $|\psi_2) \in \mathcal{H}_q$ and their coherent state representations, it is not difficult to prove the self-adjointness property of both $X_\theta^L$ and $T_\theta^L$, w.r.t. the inner product \eqref{innpro_voros}. One thing to note here is that since this analysis will not involve any integration by parts i.e. it will not involve the integration measure, this self-adjointness property of $X_\theta^L$ and $T_\theta^L$ will continue to hold for the `induced' inner product \eqref{5a5} as well.

The corresponding expressions for right action are obtained as
\begin{equation}
X_\theta^R \equiv \left[x+\frac{\theta}{2}(\partial_x+i\partial_t)\right];~T_\theta^R \equiv \left[t+\frac{\theta}{2}(\partial_t-i\partial_x)\right].\label{49}
\end{equation}
Finally noting that the adjoint action of momenta operators in (\ref{20}) are essentially given by the difference of the left and right actions and this allows us to write
\begin{equation} \label{mom_op_act}
\hat{P}_t \psi(\hat{x},\hat{t})=\frac{1}{\theta}(\hat{X}_L - \hat{X}_R)\psi(\hat{x},\hat{t});~ \hat{P}_x \psi(\hat{x},\hat{t})=-\frac{1}{\theta}(\hat{T}_L - \hat{T}_R)\psi(\hat{x},\hat{t}).
\end{equation}
We then take the overlap with $|x,t)$ to yield
\begin{equation}
(x,t|\hat{P}_t \psi(\hat{x},\hat{t}))= -i \partial_t\psi(x,t)~;~~ (x,t|\hat{P}_x \psi(\hat{x},\hat{t}))=-i\partial_x\psi(x,t)
\end{equation}
where we have made use of (\ref{47}-\ref{49}). Thus, unlike the space-time operators $\left(T_\theta, X_\theta\right)$, $\hat{P}_t$ and $\hat{P}_x$ retain their commutative form \eqref{n2} :
\begin{equation} \label{rep_p}
\hat{P}_t = -i \partial_t ~,~ \hat{P}_x = -i \partial_x
\end{equation}
We can now introduce commuting time ($\hat{T}_c$) and space ($\hat{X}_c$) variables \cite{Bal3} by taking the average of left and right actions\footnote{As an aside, we would like to mention that common eigenstates $|x,t)_M$ satisfying $\hat{X}_c |x,t)_M = x |x,t)_M$ and $\hat{T}_c |x,t)_M = t |x,t)_M$ can now be easily constructed as in \cite{Basu} and can be identified as Moyal basis, as the counterpart of \eqref{b7} can also be written, where the corresponding symbols compose the Moyal star product $\star_M$ \eqref{b8}}(see appendix \ref{app1} for further discussions)
\begin{equation} \label{comm_op}
\hat{T}_c = \frac{\hat{T}_L + \hat{T}_R}{2} ~~\mathrm{and}~~ \hat{X}_c = \frac{\hat{X}_L + \hat{X}_R}{2} ~;~ \left[ \hat{T}_c, \hat{X}_c \right] = 0
\end{equation}
We can then write the effective commutative Schr\"odinger equation by taking the overlap of \eqref{sc} with $|x, t )$ and making use of \eqref{b7} as,
\begin{equation} \label{eff_se}
i \frac{\partial \psi}{\partial t} = - \frac{1}{2m} \frac{\partial^2 \psi}{\partial x^2} + V(x,t) \star \psi (x,t)
\end{equation}
This can alternatively be written as
\begin{equation} \label{eff_se_2}
i \frac{\partial \psi}{\partial t} = - \frac{1}{2m} \frac{\partial^2 \psi}{\partial x^2} + V(X_\theta,T_\theta) \psi (x,t)
\end{equation}
Thus the effect of noncommutativity ``sneaks" in through two avenues : one through the coherent state basis $|x,t)$ and the other through the Voros star product. The second one, of course occurs only in presence of potential $V(x,t)$. In view of the presence of infinite order of derivatives in \eqref{b8}, the effective commutative theory therefore becomes non-local.

In terms of the induced inner product \eqref{5a5} the continuity equation \eqref{29} now takes the following form
\begin{equation} \label{cont_eq}
\partial_t \rho (x,t) + \partial_x j(x) = 0
\end{equation}
with the probability density $\rho(x)$ is given by \eqref{b6} and current density $j(x)$ has the following form in coherent state basis :
\begin{equation} \label{eqa5}
j (x) = \frac{i}{2m} \left( \psi^\ast (x,t) \star \frac{\partial \psi (x,t)}{\partial x} - \frac{\partial \psi^\ast (x,t)}{\partial x} \star \psi (x,t) \right)
\end{equation}
In view of the positivity condition \eqref{eq40} $\rho(x,t)$ can indeed be interpreted as probability density for a particular time - as mentioned earlier. The total probability at a time $t$, which should be normalized to $1$, is to be obtained by integrating over only $x$ and using the ``induced" inner product \eqref{5a4}. \
We can now check that $\psi_E (x,t)$ \eqref{eq50} satisfies the following effective time-independent Schr\"odinger equation for time-independent potential $V(x)$ :
\begin{equation} \label{eqb2}
E \, \psi_E (x, t) = - \frac{1}{2m} \frac{\partial^2 \psi_E (x,t)}{\partial x^2} ~+~ V(x) \star \psi_E (x,t)
\end{equation}

With all these formal aspects of our formalism in place, we can now study its application to some quantum mechanical systems. In the next section, we begin with an analysis of free particle and later we study the behaviour of a particle under the harmonic potential.


\section{Free particle wave packet in the noncommutative space-time} \label{sec5}

In this section we intend to construct a Gaussian wave packet for a free particle moving under the hamiltonian,
\begin{equation} \label{H_free}
\hat{H} = \frac{\hat{P}^2}{2m}
\end{equation}
and study the possible signatures of noncommutativity that can be observed in its time evolution.
Let us first consider an operator $\hat{\rho}$ defined as
\begin{equation} 
\hat{\rho} = \int dp dE ~\delta(E - E_p) \, |p,E)(p,E| = \int dp \, |p,E_p)(p,E_p| ~;~ E_p=\frac{p^2}{2m}
\end{equation}
Here we have inserted an appropriate delta function in the completeness relation \eqref{24}, implementing non-relativistic ``on-shell" condition for each momentum component. Its action on a generic state $|\psi)$, such as $|\psi) = \int dp dE ~| p,E ) ( p,E |\Psi ) \in \mathcal{H}_q$ is then given by,
\begin{equation} \label{rho_action}
\hat{\rho} |\psi)~\equiv \int dp ~ \psi(p,E_p) ~ |p,E_p) ~~;~ \psi(p,E_p) = ( p,E_p |\Psi )
\end{equation}
We can now write, for the inner product $( p', E_{p'} | p, E_p )$ by inserting the identity operator \eqref{idn} to get
\begin{equation}
\left( p^\prime, E_{p^\prime} | p, E_p \right) = \int dt dx \, ( p^\prime, E_{p^\prime} | x,t ) \star ( x,t | p,E_p ) = \int dt \, \left( p^\prime, E_{p^\prime} | \pi_t | p, E_p \right) = \frac{1}{2\pi} \delta(p^\prime - p) \int dt
\end{equation}
where we have made use of \eqref{pi_t}.
As before, this too diverges. Again since time `$t$' is now excluded from configuration space, we can extract the usual quantum mechanical inner product, as in \eqref{5a4}, where only a spatial integration over a constant $t$-surface occurs :
\begin{equation} \label{7a1}
\left( p^\prime, E_{p^\prime} | p, E_p \right)_t = \int_t dx \, ( p^\prime, E_{p^\prime} | x,t ) \star ( x,t | p,E_p ) = \frac{1}{2\pi} \delta(p^\prime - p)
\end{equation}
In some generalised sense, the above operator $\hat{\rho}$ \eqref{rho_action} too satisfies the property of a projection operator :
\begin{equation} \label{7a3}
\hat{\rho} \, \pi_t \, \hat{\rho} = \hat{\rho}
\end{equation}
when the ``quasi-projection operator" $\pi_t$ (\ref{pi_t}, \ref{7a2}) is sandwiched between $\hat{\rho}$'s. We now consider a Gaussian wave function for the free particle in momentum space as :
\begin{equation} \label{gwp_mom}
( p,E_p | \psi )~\equiv\psi(p,E_p) = \frac{\sqrt{\sigma}}{\pi^{1/4}} \, e^{-\frac{\sigma^2 p^2}{2}}
\end{equation}
The coherent state representation of $\hat{\rho} |\psi)$ \eqref{rho_action} corresponding to \eqref{gwp_mom} then yields
\begin{equation} \label{gwp}
\begin{split}
\Psi (x,t) = \left( x,t | \hat{\rho} | \psi \right) = \frac{\sigma^{1/2}}{2\pi^{5/4}} \, \int dp \, e^{-\frac{\theta p^4}{16 m^2} - \lambda p^2 + ipx} ~~;~ \lambda = \left( \frac{\sigma^2}{2} + \frac{\theta}{4} + i \frac{t}{2m} \right) ~~\mathrm{is~a~constant}
\end{split}
\end{equation}
Retaining terms upto first order in $\theta$, one can then show that $\Psi (x,t)$ can be recast in the following form,
\begin{equation}
\Psi (x, t) \simeq \frac{1}{2\pi^{3/4}} \sqrt{\frac{\sigma}{\lambda}} \,\, \left[ 1 + \theta \, f(x; \lambda) \right] e^{-\frac{x^2}{4\lambda}}
\end{equation}
where, the function $f(x;\lambda)$ is
\begin{equation*}
f(x; \lambda) = \frac{1}{16m^2} \left( - \frac{3}{4\lambda^2} + \frac{3x^2}{4\lambda^3} - \frac{x^4}{16\lambda^4} \right)
\end{equation*}
displaying a slight deviation in the functional form, away from Gaussian one in coordinate space. The $\theta$-deformation comes in the exponential term as well as in the amplitude also. The width $d$ of the deformed Gaussian term at time $t$ is now found to get enhanced due to noncommutativity :
\begin{equation} \label{d}
d = \sqrt{2 \abs{\lambda}} = \left[ \left( \sigma^2 + \frac{\theta}{2} \right)^2 + \left( \frac{t}{m} \right)^2 \right]^{\frac{1}{4}}
\end{equation}
This shows that even for an infinite spread in the Gaussian wave function \eqref{gwp_mom} in the momentum space ($\sigma \to 0$), the spread in coordinate space $x$ can not be squeezed below $\sim \sqrt{\frac{\theta}{2}}$.

\section{Schr\"odinger equation and energy spectra of harmonic oscillator} \label{sec6}

In this section, we start with writing the operatorial form of the Schr\"odinger equation \eqref{sc} for the time-independent harmonic oscillator potential $V(\hat{X}) = \frac{1}{2} m \omega^2 \hat{X}^2$ :
\begin{equation}
[\hat{x},\hat{\psi}] = -\frac{1}{2m\theta}[\hat{t},[\hat{t},\hat{\psi}]] -\frac{\theta}{2}m\omega^2\hat{x}^2\hat{\psi}.\label{sch}
\end{equation}
On substituting the most general form of the ansatz for the abstract state $\hat{\psi} (\hat{x}, \hat{t})$
\begin{equation}
\hat{\psi}(\hat{x},\hat{t}) = \int dE dp \, e^{-i(E\hat{t} - p\hat{x})} \tilde{\psi}(E,p),
\end{equation}
in the above equation \eqref{sch} yields the following time-independent form of Schr\"odinger equation in energy-momentum space :
\begin{equation} \label{ll}
\frac{1}{2m} \left[p^2 - m^2 \omega^2\left(\frac{\partial}{\partial p} + \frac{i}{2}\theta E \right)^2\right] \tilde{\psi} (E,p) = E\tilde{\psi} (E,p)
\end{equation}
Introducing creation and annihilation operators as
\begin{equation}
a_E = \frac{1}{\sqrt{2m\omega}}\left[p_x + m\omega \left(\frac{\partial}{\partial p_x}+\frac{i\theta E}{2}\right)\right] ~;~ a_E^\dagger = \frac{1}{\sqrt{2m\omega}} \left[p_x - m\omega \left(\frac{\partial}{\partial p_x}+\frac{i\theta E}{2}\right)\right] ~;~ [a_E, a_E^\dagger] = 1
\end{equation}
we can re-write the above equation as
\begin{equation}
\omega\left(a_E^\dagger a_E + \frac{1}{2}\right) \tilde{\psi}=E\tilde{\psi}
\end{equation}

As a noncommutative effect, one can see that $E$ occurs on both sides of this equation. However, as it turns out that this is not a serious hurdle, as this dependence of $E$ can be removed easily by energy-momentum dependent $U(1)$ transformation. To show this explicitly, let us begin by considering $\tilde{\psi_0} (p)$, the stationary state wave function corresponding to the ground state with energy $E_0$ in the momentum space. Then requiring $a_E \tilde{\psi_0} (p) = 0$, one finds the \emph{un-normalized} wave function factorises as,
\begin{equation}
\tilde{\psi_0} (p) = e^{-\frac{i\theta}{2} E_0 p} \,\, \psi_0(p) ~~;~ \psi_0 (p) = e^{-\frac{p^2}{2m\omega}}
\end{equation}
From here, we can easily show that
\begin{equation}
\left(a_E^\dagger a_E + \frac{1}{2}\right) \, e^{-\frac{i\theta}{2} E_0p} \, \, \psi_0 (p) = e^{-\frac{i\theta}{2} E_0p} \, \left(a^\dagger a + \frac{1}{2}\right) \psi_0(p)
\end{equation}
Here $a$ is the undeformed annihilation operator and $\psi_0(p)$ is the undeformed solution. This can be generalised easily to higher energy levels with undeformed wave functions $\psi_n(p)$ satisfying
\begin{equation} \label{7c1}
\left(a_E^\dagger a_E + \frac{1}{2}\right) \, e^{-\frac{i\theta}{2} E_np} \, \, \psi_n (p) = e^{-\frac{i\theta}{2} E_np} \, \left(a^\dagger a + \frac{1}{2}\right) \psi_n(p)
\end{equation}
which are related to the deformed ones as
\begin{equation} \label{7c2}
\tilde{\psi}_n (p) = e^{-\frac{i\theta}{2} E_n p} \, \, \psi_n (p)
\end{equation}
with $E_n = \left( n + \frac{1}{2} \right) \omega$. This shows that the energy spectra of the harmonic oscillator will not be deformed due to the noncommutativity of the space-time. This corroborates the observation made in \cite{Bal}. However, the corresponding wave function in coordinate space, i.e. in the coherent state can be easily obtained by making use of \eqref{eq50} and \eqref{7c2} to get the following form of \emph{un-normalised} ground state wave function with respect to the inner product (\ref{5a4}, \ref{5a5}) :
\begin{equation} \label{8a1}
\tilde{\psi_0}(x,t) = e^{-\left[\frac{(x-\frac{\theta E_0}{2})^2}{2\sigma^2_\theta}\right]} \, e^{-iE_0 t} ~~;~ \sigma^2_\theta = \frac{\theta}{2}+\frac{1}{m \omega}
\end{equation}
displays a parity violating shift in the origin and a modified width $\sigma_\theta$. Clearly, this resembles the form of the ground state wave function in commutative quantum mechanics, except for the $\theta$-deformation in the width $\sigma_\theta$. With $\theta \to 0$, one gets back the familiar commutative form of the \emph{un-normalised} ground state wave function.
\begin{equation}
\lim\limits_{\theta \to 0} \tilde{\psi_0} (x,t) = e^{- \frac{x^2}{2\sigma_0^2} } \, e^{-iE_0 t}
\end{equation}
Now coming back to \eqref{8a1}, one needs to normalise it with respect to the `induced' inner product \eqref{5a5} i.e. compute the probability density $\rho$ using \eqref{b6} and integrate it over $x$ to set it to one. Thus, with the correct normalization factor, we get
\begin{equation} \label{gr_ho_pd}
\rho(x) = \frac{1}{\tilde{\sigma_\theta} \sqrt{2\pi}} e^{-\left[\frac{(x-\theta E_0)^2}{2\tilde{\sigma_\theta}^2}\right]} ~;~ \tilde{\sigma_\theta}^2 = \frac{\sigma_\theta^2}{2} \left[ 1 + \frac{\theta}{2\sigma^2_\theta} \right]
\end{equation}
One can also re-cast this, using \eqref{a4} as,
\begin{equation} \label{gr_ho}
\rho(x) = \delta_{\tilde{\sigma_\theta}} \left( x - \theta E_0 \right)
\end{equation}
which is manifestly normalized as have been shown in \eqref{a4}. Note that the time-dependent factors in \eqref{8a1} cancels out in \eqref{gr_ho}, thus making the probability density independent of time. We can now study the effect of infinitely large confining potential, by considering the limit $\omega \to \infty$ in presence of noncommutativity $(\theta \neq 0)$. In this limit, $\tilde{\sigma_\theta} \to \sqrt{\theta}$, preventing the squeezing of the packet in a region $\lesssim \sqrt{\theta}$. This is a purely a noncommutative effect as the inherent noncommutativity in space-time provides an impenetrable barrier and does not allow for a localization to a point. Thus the probability density $\rho(x)$ at some point $x$ will have contributions from the vicinity and points from finite, however small, region around that point. In a certain sense therefore, noncommutativity thus essentially introduces a non-locality in the theory, where the notion of a point particle, it seems, has to be necessarily replaced by some extended object of a `cloud' A similar situation was observed also in the context of spatial noncommutativity \cite{RS}.

Again if one takes both the limit $\omega \to \infty$ as well as $\theta \to 0$, holding $\omega\sqrt{\theta}$, a dimensionless constant, fixed then $\rho(x) \to \delta(x)$, as one can expect in the commutative limit. One should note that the the non-local features appear for both the ground state wave function \eqref{8a1} and the probability density \eqref{gr_ho}. One can thus expect the effect of noncommutativity to appear through non-local behaviours in measurable quantities.

\section{Expectation values, uncertainty relations and Ehrenfest theorem in the coherent state basis} \label{sec7}

In this section, as an illustration, we compute the expectation values of $X_\theta$, $X_\theta^2$, $T_\theta$, $T_\theta^2$, $P_x$ and $P_x^2$ in the ground state of harmonic oscillator and study the uncertainty relations after that we study the Ehrenfest theorem in the coherent state basis. The expectation value of an arbitrary operator $\mathcal{O}_\theta$ in the coherent state basis in any stationary state $\psi_n$ with energy $E_n$ is given by
\begin{equation}
\left\langle \mathcal{O}_\theta \right\rangle_t =\int dx~ \tilde{\psi}_n^\ast (x,t) \star \mathcal{O}_\theta \, \tilde{\psi}_n (x,t).
\end{equation}
The expectation value of $X_\theta$, in the ground state \eqref{8a1}, in particular, is given by
\begin{equation}
\left\langle X_\theta\right\rangle_t = \int dx~ \tilde{\psi}^{\ast}_0(x,t) \star \left[x+\frac{\theta}{2}(\partial_x-i\partial_t)\right]\tilde{\psi}_0(x,t),
\end{equation}
After straight forward calculation, we get
\begin{equation}
\left\langle X_\theta\right\rangle_t=0.\label{hr1}
\end{equation}
despite having a shift in the right by an amount $(\frac{\theta}{2}E_0)$ in the wave function $\psi_0 (x,t)$ \eqref{8a1}. In a similar fashion, we find that
\begin{equation}
\left\langle T_\theta\right\rangle_t=t;~\left\langle T_\theta^2\right\rangle_t=t^2 +\frac{\theta}{2}+\frac{\theta^2 m\omega}{2} ;~\left\langle X_\theta^2\right\rangle_t =\frac{1}{2m\omega} \label{hr2}
\end{equation}
and
\begin{equation}
\left\langle P_{x}\right\rangle_t=0 ;~\left\langle P_x^2\right\rangle_t=\frac{m\omega}{2} . \label{hr3}
\end{equation}
displaying a non-trivial noncommutative deformation only in $\langle T_\theta^2 \rangle_t$, which, however, has the expected commutative limit.

\paragraph*{Uncertainty relations\\}

Now we compute the uncertainties in the measurement of all these observables of the harmonic oscillator. Again we work in the ground state of the harmonic oscillator. By making use of the form of the standard deviation of an observable, defined as
\begin{equation}
\Delta\mathcal{O}_\theta= \sqrt{\left\langle {\mathcal{O}_\theta}^2 \right\rangle -\left\langle \mathcal{O}_\theta \right\rangle^2}\label{hr4}
\end{equation}
and using (\ref{hr1},\ref{hr2},\ref{hr3}) and (\ref{hr4}), we get
\begin{equation}
\Delta X_\theta= \sqrt{1/(2m\omega)}~; ~~~\Delta T_\theta=\sqrt{\frac{\theta}{2}+\frac{\theta^2 m\omega}{2}};~~~\Delta P_x=\sqrt{ \frac{m\omega}{2}}
\end{equation}
Thus we arrive at the following uncertainty product relations
\begin{equation}
\Delta X_\theta\Delta T_\theta= \frac{\theta}{2}\sqrt{1+\frac{1}{m\omega\theta}} ~~;~ \Delta X_\theta\Delta P_x= \frac{1}{2}
\end{equation}
Note that for infinitely confining potential $(\omega \to \infty)$, $\Delta X_\theta \to 0$, $\Delta T_\theta \to \infty$ but their product $\Delta X_\theta \Delta T_\theta \to \frac{1}{2}$, as expected. It is worth mentioning at this point that the coherent states correspond to the saturation of uncertainty relations. Now the ground state of the harmonic oscillator is a coherent state in the phase space and therefore the uncertainty relation between position and momentum is saturated. On the other hand, the same state can not be identified as the one stemming from coherent state for the configuration space i.e. $\mathcal{H}_c$ \eqref{b1} and so it is not a great surprise that the position-time uncertainty is not saturated. Indeed we shall show by explicit computation now that position-time uncertainty is minimum i.e. saturated if one performs a similar calculation in the coherent state $|z)$ \eqref{vbasis} in $\mathcal{H}_q$ instead of energy eigenstates in the phase space. Note that for coherent state $\ket{z} \in \mathcal{H}_c$ \eqref{b3}, this is already known to satisfied, as mentioned earlier.

Finally, note that in this ground state $\Delta E = 0$, consequently $\Delta E \Delta T_\theta$ also vanishes and this is compatible with the uncertainty relations $\Delta \hat{H} \Delta \hat{T} \geq \frac{1}{2} m \omega^2 \theta \abs{\langle \hat{X}_\theta \rangle_t}$, as follows from \eqref{hr1}. One can not thus expect the energy-time uncertainty to arise from this kind of noncommutative structure.

\paragraph*{Generalized Schr\"odinger Uncertainty Relation \\}

In this context, let us check the status of the general Schr\"odinger-Robertson uncertainty relation for the coherent state basis \eqref{vbasis} by constructing the variance matrix $V^\theta$ for the operators $\left\lbrace \hat{X}, \hat{T}, \hat{P}_x, \hat{P}_t \right\rbrace$. The essential construction of such uncertainty relation has been presented in the appendix. We take up from there and start by using (\ref{19}, \ref{20}). It is then straightforward to compute that
\begin{equation} \label{vmatrix}
V^\theta = \begin{pmatrix}
\theta/2 & 0 & 0 & -1/2\\
0 & \theta/2 & 1/2 & 0 \\
0 & 1/2 & 1/\theta & 0 \\
-1/2 & 0 & 0 & 1/\theta \
\end{pmatrix}
\end{equation}
where the rows represent the operators $\hat{X}, \hat{T}, \hat{P}_x, \hat{P}_t$ respectively from the left and the columns represent the same set of operators from the top in the same order.

The determinant of the variance matrix gives us the uncertainty product. Here
\begin{equation}
Det \, V^\theta = \frac{3}{16} > \frac{1}{4^2}
\end{equation}
which implies that for the whole phase space the basis \eqref{vbasis} does not have minimum uncertainty product. Rather from the structure of \eqref{vmatrix}, one can see that if we consider only the configuration space $Span \, \left\lbrace \hat{X}, \hat{T} \right\rbrace$ then the uncertainty product is
\begin{equation}
\Delta \hat{X} \Delta \hat{T} = \sqrt{Det \, \left(V^\theta\right)_{2 \times 2}} = \frac{\theta}{2}
\end{equation}
which is well understood from the fact that \eqref{vbasis} is actually a coherent state with respect to the configuration space variables (see \eqref{b4}).

\paragraph*{Ehrenfest theorem in the coherent state basis \\}
As we have shown in the section 2 that the primary constraint annihilate the physical state then we can write
\begin{equation}
(\psi|[\hat{H}, \hat{\mathcal{O}}]|\psi) = -(\psi|[\hat{P}_t, \hat{\mathcal{O}}]|\psi).\label{eh1}
\end{equation}
Now using completeness relation \eqref{idn} for the coherent state basis, we can write above equation as
\begin{equation}
\begin{split}
(\psi|\hat{P}_t \hat{\mathcal{O}}|\psi)-(\psi|\hat{\mathcal{O}}\hat{P}_t|\psi) & = \int dxdt ~ \left[ (\psi|\hat{P}_t|x,t) \star (x,t| \hat{\mathcal{O}}|\psi) - (\psi|\hat{\mathcal{O}}|x,t) \star (x,t|\hat{P}_t|\psi)\right] \\
 & = \int dxdt \, \left[ i\partial_t \psi^\ast (x,t) \star \mathcal{O}_\theta \psi(x,t) + \mathcal{O}^\ast_\theta \psi^\ast(x,t) \star i\partial_t \psi(x,t) \right]
\end{split}
\end{equation}
For a self-adjoint operator $\mathcal{O}_\theta$, the above equation reduces to
\begin{equation}
\begin{split}
(\psi|\hat{P}_t \hat{\mathcal{O}}|\psi)-(\psi|\hat{\mathcal{O}}\hat{P}_t|\psi)=\int dt \, \left(i\partial_t \int dx \, \psi^\ast (x,t) \star \mathcal{O}_\theta \psi(x,t) - i \int dx \, \psi^\ast(x,t) \star_V \left(\partial_t \mathcal{O}_\theta\right) \psi(x,t)\right)
\end{split}
\end{equation}
Equivalently,
\begin{equation} \label{eh2}
(\psi|\hat{P}_t \hat{\mathcal{O}}|\psi)-(\psi|\hat{\mathcal{O}}\hat{P}_t|\psi) = \int dt \, \left(i\partial_t \left\langle\mathcal{O}_\theta \right\rangle_t - i \left\langle \partial_t \mathcal{O}_\theta \right\rangle_t \right) 
\end{equation}
Now let us consider
\begin{equation} \label{eh3}
\begin{split}
(\psi|[\hat{H}, \hat{\mathcal{O}}]|\psi) & = \int dxdt \, (\psi|x,t) \star (x,t|[\hat{H}, \hat{\mathcal{O}}]|\psi) \\
 & = \int dxdt \, \psi^\ast (x,t) \star [\hat{H}, \hat{\mathcal{O}}]_\theta\psi(x,t) \\
 & = \int dt \, \left\langle[\hat{H}, \hat{\mathcal{O}}]_\theta\right\rangle_t \
\end{split}
\end{equation}
Using \eqref{eh2} and \eqref{eh3} in \eqref{eh1} we get the following undeformed version of Ehrenfest theorem
\begin{equation}
\partial_t \left\langle\mathcal{O}_\theta \right\rangle_t=i\left\langle[\hat{H}, \hat{\mathcal{O}}]_\theta\right\rangle_t + \left\langle\partial_t \mathcal{O}_\theta \right\rangle_t.
\end{equation}
although the representation of the operators get modified due to the noncommutativity in coherent state basis. Now on using above equation we can easily get, for the particular cases of various observables the following results :
\begin{equation}
\partial_t \left\langle X_\theta\right\rangle_t =\frac{\left\langle P_x\right\rangle_t}{m},
\end{equation} 
\begin{equation} \label{eq116}
\partial_t  \left\langle P_x\right\rangle_t=-\left \langle\frac{\partial V(x)}{\partial x}\right\rangle_t - \frac{\theta}{2} \left\langle\frac{\partial^2 V(x)}{\partial x^2} (\partial_x -i\partial_t)\right\rangle_t + \mathcal{O}(\theta^2),
\end{equation}
and
\begin{equation}
\partial_t \left\langle T_\theta\right\rangle_t = 1 + \theta\left\langle \frac{\partial V(x)}{\partial x}\right\rangle_t + \frac{\theta^2}{2} \left\langle\frac{\partial^2 V(x)}{\partial x^2}(\partial_x -i\partial_t)\right\rangle_t + \mathcal{O}(\theta^3).
\end{equation}
It is now a matter of trivial exercise to verify for the ground state of the harmonic oscillator that all the above three equations are satisfied (\ref{hr1},\ref{hr2},\ref{hr3}). These noncommutative deformations suggest that they induce additional forces of noncommutative origin.

\section{Transition probability in presence of space-time noncommutativity} \label{sec8}

In this section, we intend to study the modification due to noncommutativity, if any, in the transition rate of a system when it undergoes a transition from a given initial state to a particular final state. For this study let us first consider system, described by a Hamiltonian $H_0$ and satisfies time independent Schr\"odinger equation as
\begin{equation}
H_0 \ket{\phi_n}=E_n \ket{\phi_n}.
\end{equation}
These stationary eigenstates $\ket{\phi_n}$ evolve in time as
\begin{equation}
\ket{\phi_n(t)}=e^{-iE_nt}\ket{\phi_n}.
\end{equation}
Now if we switch on a time dependent perturbation say $V(\hat{t})$ then its time evolution will be described by the Hamiltonian $H=H_0 + V(\hat{t})$ satisfying Schr\"odinger equation as
\begin{equation}
H\ket{\psi(t)} = i\frac{\partial}{\partial t}\ket{\psi(t)},
\end{equation}
We work with the ansatz for the state $\ket{\psi(t)}=\sum_{n} C_n(t)\ket{\phi_n(t)}$ where the coefficients $C_n(t)$ are themselves time dependent. In our noncommutative case, the above equation can be rewritten as
\begin{equation}
H_0\ket{\psi(\hat{x},\hat{t})}+V(\hat{t})|\psi(\hat{x},\hat{t})) = -P_t|\psi(\hat{x},\hat{t})).
\end{equation}
Now in the coherent state basis \eqref{eq35}, above equation will take the form as
\begin{equation}
(x,t|H_0|\psi)+(x,t|V(\hat{t})|\psi) = i\frac{\partial}{\partial t}(x,t|\psi),
\end{equation}
or equivalently,
\begin{equation}
H_0(P_x,X_\theta) \, \psi(x,t) + V(T_\theta) \, \psi(x,t) = i \, \partial_t \psi(x,t).
\end{equation}

Let us now consider a generic non-stationary state which satisfies the above Schr\"odinger equation as $\psi(x,t) = \sum\limits_n C_n (t) e^{-iE_n t}\phi_n(x)$, where the subscript $n$ refers to their association to the energy levels $E_n$. Now substituting $\psi(x,t)$ in the above equation we get,
\begin{equation}
\begin{split}
\sum_n i\dot{C}_n (t) \, & e^{-iE_n t}\phi_n(x) + \sum_n E_n C_n (t) \, e^{-iE_n t}\phi_n(x) \\
 & = \sum_n H_0(P_x,X_\theta) \, C_n (t) \, e^{-iE_n t}\phi_n(x) + \sum_n V(T_\theta) \, C_n (t) \, e^{-iE_n t}\phi_n(x) \
\end{split}
\end{equation}
Simplifying,
\begin{equation}
\sum_n i\dot{C}_n (t) \, e^{-iE_n t}\phi_n(x) = \sum_n V(T_\theta) \, C_n (t) \, e^{-iE_n t}\phi_n(x).
\end{equation}
Now using the representation of the time operator \eqref{48} in the coherent state basis and expanding $V(T_\theta)$ by Taylor`s series upto first order in $\theta$ we find
\begin{equation}
\begin{split}
\sum_n i\dot{C}_n (t) \, e^{-iE_n t} \, \phi_n(x) & \simeq \sum_n V(t) \, C_n (t)e^{-iE_n t} \phi_n(x) \\
 & + \frac{\theta}{2}\sum_n C_n (t)\frac{\partial V(t)}{\partial t} \left( \partial_t + i\partial_x \right) e^{-iE_n t}\phi_n(x) + \mathcal{O} (\theta^2) \
\end{split}
\end{equation}
We can recast the above equation by star multiplying both sides from the left by $e^{iE_mt} \, \phi_m^{\ast} (x;E_m)$ as
\begin{equation}
\begin{split}
\sum_n i\dot{C}_n (t) \, e^{iE_m t} \, \phi_m^{\ast}(x) \star e^{-iE_n t} \, \phi_n(x) & \simeq  \sum_n V(t) \, C_n (t) \, \phi_m^{\ast}(x) \, \star \, e^{-iE_n t} \, \phi_n(x) \\
 & + \frac{\theta}{2} \sum_n C_n (t) \, \phi_m^{\ast}(x) \, \star \, \frac{\partial V(t)}{\partial t} \, (\partial_t +i\partial_x) \, e^{-iE_n t} \, \phi_n(x) \
\end{split}
\end{equation}
Here we are assuming that $|\theta \dot{C}_n|<<1$ which is understood from the fact that $C_n (E_n,t)$ contains $\theta$ factor as we have discussed in \eqref{eqb3}. Then from the above equation we can easily get
\begin{equation}
\begin{split}
\sum_n i\dot{C}_n (t) \, \delta_{E_m,E_n} & \simeq \sum_n C_n (t) \, \left\langle \phi_m |V(t)|\phi_n \right\rangle + \frac{\theta}{2} \sum_n C_n (t) \, \langle \phi_m |\frac{\partial V(t)}{\partial t}(\partial_t +i\partial_x) | \phi_n \rangle \
\end{split}
\end{equation}
where $\phi_n (x,t) = e^{-iE_n t}\phi_n(x)$. If the time dependence perturbation is too small then one can neglect the $C_n (t)$ for $n \neq i$ as this develop only because of the perturbation and we require only $C_i (t)$ but for weak perturbation $V(t)$ the coefficient $C_i (t)$ are approximately same as $C_i (t=0)$ then
\begin{equation}
i\dot{C}_m (t) \simeq C_i (t=0) \, \langle \phi_m |V(t)|\phi_i \rangle + \frac{\theta}{2} C_i (t=0) \langle \phi_m |\frac{\partial V(t)}{\partial t}(\partial_t +i\partial_x)| \phi_i \rangle
\end{equation}
Therefore if the system undergoes transition in time $T$, then for initial state $| i \rangle$ and final state $| f \rangle$, we have
\begin{equation}
\frac{C_f (E_f,T)}{C_i (E_i,t=0)} \simeq -i \int_0^T dt \, \left[ \langle f |V(t)| i \rangle + \frac{\theta}{2} \, \langle f| \frac{\partial V(t)}{\partial t}(\partial_t +i\partial_x) | i \rangle \right]
\end{equation}
Now the relative probability of transition from an initial state $\ket{i}$ to a final state $\ket{f}$ is defined as,
\begin{equation}
P_{i \to f} = \abs{\frac{C_f (E_f,T)}{C_i (E_i,t=0)}}^2
\end{equation}
The relative transition rate of the system for a total time $T$ is then given by
\begin{equation}
\begin{split}
T_{i \to f} & = \frac{P_{i \to f}}{T} = \frac{1}{T} \, \abs{\int_0^T dt \, \left[ \langle f |V(t)|i \rangle + \frac{\theta}{2} \, \langle f| \frac{\partial V(t)}{\partial t}(\partial_t + i\partial_x) | i \rangle \right]}^2 \
\end{split}
\end{equation}
 
In the limit $\theta \rightarrow 0$, we get back the commutative result. The presence of $\theta$ dependent term in the transition rate is clearly a noncommutative effect for which this rate is found to get enhanced.

\section{Galilean algebra and Galilean generators in Moyal space-time} \label{sec9}

In this section we study the Galilean generator and Galilean algebra for Moyal space-time. Let us consider the particle is moving in the $x$-axis then the Galilean transformation is given as
\begin{equation}
\hat{x} \longrightarrow \hat{x}^\prime = \hat{x}-v\hat{t} ~;~ \hat{t} \longrightarrow \hat{t}^\prime = \hat{t} 
\end{equation}

The Schr\"odinger equation for a free particle in the prime coordinates frame is given as
\begin{equation}
[\hat{x}^\prime,\psi^\prime]=-\frac{1}{2m\theta}[\hat{t}^\prime,[\hat{t}^\prime,\psi^\prime]],
\end{equation}
where $\psi\longrightarrow\psi^\prime =\hat{U}\psi$ with $\hat{U}$ being a unitary operator, to be determined. Now using prime coordinates in the terms of unprimed coordinates in above equation, we get
\begin{equation}
\left([\hat{x},\hat{U}]+\frac{1}{2m\theta}[\hat{t},[\hat{t},\hat{U}]]\right)\psi -v\hat{U}[\hat{t},\psi]-v[\hat{t},\hat{U}]\psi=-\frac{1}{m\theta}[\hat{t},\hat{U}][\hat{t},\psi].
\end{equation}
Let us assume $\hat{U}=e^{iv\hat{\phi}(\hat{x}, \hat{t})}=1+iv\hat{\phi}(\hat{x}, \hat{t})$ (up to first order in $v$ with $\abs{v} \ll 1$) and use it in the above equation. We then find
\begin{equation}
-\frac{i\theta}{2m}\hat{\phi}^{\prime\prime}-[\hat{t},\psi]=\frac{1}{m}\phi^\prime[\hat{t},\phi],
\end{equation}
where $\hat{\phi}^\prime = \frac{d\hat{\phi}}{d\hat{x}} := -\frac{i}{\theta} \left[ \hat{t}, \hat{\phi} \right]$. Now equating the coefficients of $[\hat{t},\psi]$ from both sides in the above equation, we get
\begin{equation}
\hat{\phi}(\hat{x})=-m\hat{x}.
\end{equation}
Also, we can write the Galilean transformation in the matrix form as
\begin{equation}
\begin{pmatrix}
\hat{x}^\prime\\ \hat{t}^\prime
\end{pmatrix}
=\begin{pmatrix}
1 & -v\\0 & 1
\end{pmatrix}
\begin{pmatrix}
\hat{x}\\\hat{t}
\end{pmatrix},
\end{equation}
where the Galilean boost matrix $B$ is given by $B = \begin{pmatrix}
1 & -v\\0 & 1
\end{pmatrix}$ and we know that $\psi^\prime(\hat{x})=\psi(B^{-1}\hat{x})$. Then 
\begin{equation}
\psi^\prime(\hat{x},\hat{t}) = e^{-imv(\hat{x}+v\hat{t})} \psi(\hat{x}+v\hat{t},\hat{t})
\end{equation}
The solution of the free particle Schr\"odinger equation is $\psi(\hat{x},\hat{t}) = e^{-i(E \hat{t}-p\hat{x})}$ and use this in the above equation, we find
\begin{equation}
\psi^\prime(\hat{x},\hat{t})=e^{-imv(\hat{x}+v\hat{t})}e^{-i(E \hat{t}-p(\hat{x}+v\hat{t}))}
\end{equation}
and now on using Baker-Campbell-Hausdorff formula, we get
\begin{equation}
\psi^\prime(\hat{x},\hat{t})=e^{-imv(\hat{x}+v\hat{t})}e^{-i(E \hat{t}-p\hat{x})}e^{iv\left(p\hat{t}+\frac{\theta p^2}{2}\right)}
\end{equation}
or
\begin{equation} \label{G}
\psi^\prime(\hat{x},\hat{t})=e^{-iv\hat{G}}\psi(\hat{x},\hat{t}) ~;~ \hat{G}=m\hat{X}-\hat{P}\hat{T}-\frac{\theta}{2}\hat{P}^2
\end{equation}
is deformed Galilean boost generator. The algebra satisfied by this modified generator is the following :
\begin{equation}
[\hat{G},H]=i\hat{P},~~[\hat{G},\hat{P}]=im,~~~[\hat{G},\hat{P}_t]=-i\hat{P}
\end{equation}

We conclude this section with the observation that the Galilean algebra has the same form as in the commutative case but here generators of the algebra get modified due to the noncommutativity of the space-time. Nevertheless, one can restore the commutative form by re-writing $\hat{G}$ \eqref{G} in terms of the commuting time operator $\hat{T}_c$ as
\begin{equation} \label{12a1}
\hat{G} = m\hat{X} - \hat{P}\hat{T}_c ~;~ \hat{T}_c = \hat{T} + \frac{\theta}{2} \hat{P}
\end{equation}
where we have eliminated $\hat{T}_R$ by making use of (\ref{mom_op_act}, \ref{comm_op}).

\section{Conclusion} \label{sec10}

Here we have considered a simple $(1+1)$-dimensional quantum mechanics where both spatial and temporal coordinates are operator-valued and satisfy the simplest type of noncommutative algebra \eqref{intro2}. We show that by making use of coherent state basis \eqref{vbasis} it is possible to write down an effective commutative theory starting from an abstract version of Schr\"odinger equation \eqref{eff_se} using Hilbert-Schmidt operators. This equation now enjoys almost a similar form to that of commutative quantum mechanics except that the point-wise product of functions are necessarily replaced by Voros star product; a Moyal star product could not be useful as this fails to ensure positive definiteness of probability density. Also since the star products involve terms with infinite order derivatives of spatio-temporal variables, this theory becomes essentially non-local. Also, since we start from a time-reparametrisation invariant form of an action, both space and time coordinates enjoy similar status, in the sense that they are both members of configuration space and eventually are promoted to the level of operators. Naturally, the associated Hilbert space is `bigger' and we needed to introduce the so-called `induced' inner product so that it lands itself to the conventional probabilistic interpretation.

We then study various applications from the basic formalism described so far and consider the examples of free Gaussian wave packet, harmonic oscillator. For free particle we show explicitly that, the inherent noncommutativity prevents the particle to get localized to a single point even if one introduces an infinite uncertainty in the momentum space. We then compute harmonic oscillator spectra and we find that the spectra remains undeformed, although the corresponding wave functions get deformed. These ground state wave functions displays a parity violating shift in the origin and a deformed width. We also show that for a infinitely large confining potential, one can not squeeze the position of a particle under a certain limit. This is certainly a non-local feature of the theory arising from the space-time noncommutativity. We then obtain noncommutativity induced deformation in various fundamental uncertainty relations and we propose a noncommutative modification to the Ehrenfest theorem. We compute the transition probability of a particle under a time-dependent potential and find a similar deformation in the Fermi's golden rule. Finally we construct the modified Gailean generators for the noncommutative system, where the Galilean algebra retains the form as in the commutative case.

Presently, we are working to extend our formulation to study second quantized theories in noncommutative space-time.

\section*{Acknowledgements}

One of the authors, SKP will like to thank UGC-India for providing financial support in the form of fellowship during the course of this work.

\appendix

\section{Appendix}

\subsection{Commuting space-time coordinate operators} \label{app1}

One can again make use of \eqref{mom_op_act} and \eqref{comm_op} to eliminate $\hat{X}_R$ and relate $\hat{X}_c$ and $\hat{X}$, just like \eqref{12a1}, to get
\begin{equation} \label{nc_to_c}
\begin{split}
\hat{T}_c & = \hat{T} + \frac{\theta}{2} \hat{P}_x ~;~ \hat{X}_c = \hat{X} - \frac{\theta}{2} \hat{P}_t \\
\end{split}
\end{equation}
These coordinates are clearly unphysical, as they fail to capture the noncommutative features \cite{Basu}. But one can represent the noncommutating equations in terms of these commuting operators and then it takes a form similar to its analogue in commutative quantum mechanics. In fact, in section \ref{sec9}, we have shown that in terms of $\hat{X}_c$ and $\hat{T}_c$, the Galilean generators \eqref{12a1} retains its commutative form.

Also, \eqref{nc_to_c} establishes a relation between the commuting operators and the noncommutative operators. This essentially gives us the transformation matrix $M$ connecting a commutative phase space operators $\left( \hat{Z}^c_1, \hat{Z}^c_2, \hat{Z}^c_3, \hat{Z}^c_4 \right)^T = \left( \hat{T}_c, \hat{X}_c, \hat{P}_t, \hat{P}_x \right)^T$ and its noncommutative counterparts $\left( \hat{Z}_1, \hat{Z}_2, \hat{Z}_3, \hat{Z}_4 \right)^T$
\begin{equation} \label{M}
\hat{Z}^c_\mu = M_{\mu\nu} \hat{Z}_\nu ~;~ M = \begin{pmatrix}
1 & 0 & 0 & \frac{\theta}{2} \\
0 & 1 & -\frac{\theta}{2} & 0 \\
0 & 0 & 1 & 0 \\
0 & 0 & 0 & 1 \
\end{pmatrix} ~;~ det \, M = 1
\end{equation}
where we have augmented the pair of equations in \eqref{nc_to_c} by
\begin{equation}
\hat{P}^c_t  = \hat{P}_t  ~ ;~ \hat{P}^c_x  = \hat{P}_x
\end{equation}
For obvious reasons, it is not a canonical transformation.

\subsection{Generalized Robertson and Schr\"odinger uncertainty relation} \label{app2}

In this context, let us digress for a little while and discuss the status of generalized Schr\"odinger uncertainty relation for the phase space operators $\left\lbrace \hat{X}, \hat{T}, \hat{P}_x, \hat{P}_t \right\rbrace$ in the context of commutative quantum mechanics $(\theta = 0)$. We essentially follow the approach of \cite{Chao}.

The variance for any hermitian operator $\hat{\mathcal{O}}$ in a general state $| \Psi \rangle$ is given by
\begin{equation}
\left( \Delta \hat{\mathcal{O}} \right)^2 = \langle \Psi | \left( \Delta \hat{\mathcal{O}} - \left\langle \hat{\mathcal{O}} \right\rangle \right)^2 | \Psi \rangle = \left\langle f_{\mathcal{O}} | f_\mathcal{O} \right\rangle  ~~~\mathrm{with}~~ | f_\mathcal{O} \rangle = | \left( \Delta \hat{\mathcal{O}} - \left\langle \hat{\mathcal{O}} \right\rangle \right) \Psi \rangle
\end{equation}
Then, for two such observables $\hat{\mathcal{O}}$ and $\hat{\mathcal{O}}^\prime$ we have
\begin{equation}
\left( \Delta \hat{\mathcal{O}} \right)^2 \left( \Delta \hat{\mathcal{O}}^\prime \right)^2 = \langle f_\mathcal{O} | f_\mathcal{O} \rangle \langle f_{\mathcal{O}^\prime} | f_{\mathcal{O}^\prime} \rangle \geq \abs{\left\langle f_\mathcal{O} | f_{\mathcal{O}^\prime} \right\rangle }^2
\end{equation}
One can now split the term $\abs{\left\langle f_\mathcal{O} | f_{\mathcal{O}^\prime} \right\rangle }^2$ into its real and imaginary part as,
\begin{equation}
\abs{\left\langle f_\mathcal{O} | f_{\mathcal{O}^\prime} \right\rangle }^2 = \left(\frac{\left\langle f_\mathcal{O} | f_{\mathcal{O}^\prime} \right\rangle + \left\langle f_{\mathcal{O}^\prime} | f_{\mathcal{O}} \right\rangle}{2} \right)^2 + \left(\frac{\left\langle f_\mathcal{O} | f_{\mathcal{O}^\prime} \right\rangle - \left\langle f_{\mathcal{O}^\prime} | f_{\mathcal{O}} \right\rangle}{2i} \right)^2
\end{equation}

\begin{itemize}
\item For Robertson uncertainty relation, we ignore the square of the real part. Thus
\begin{equation} \label{rob_un_rel}
\Delta \hat{\mathcal{O}} \Delta \hat{\mathcal{O}}^\prime \geq \sqrt{\left(\frac{\left\langle f_\mathcal{O} | f_{\mathcal{O}^\prime} \right\rangle - \left\langle f_{\mathcal{O}^\prime} | f_{\mathcal{O}} \right\rangle}{2i} \right)^2}
\end{equation}
which gives us
\begin{equation} \label{rob_un_rel_2}
\Delta \hat{\mathcal{O}} \Delta \hat{\mathcal{O}}^\prime \geq \frac{1}{2i} \left\langle \left[ \hat{\mathcal{O}}, \hat{\mathcal{O}}^\prime \right] \right\rangle 
\end{equation}

\item Schr\"odinger uncertainty relation is obtained by retaining both the real and imaginary part,
\begin{equation} \label{sc_un_rel}
\Delta \hat{\mathcal{O}} \Delta \hat{\mathcal{O}}^\prime \geq \sqrt{\left( \frac{1}{2} \left\langle \left\lbrace \hat{\mathcal{O}}, \hat{\mathcal{O}}^\prime \right\rbrace \right\rangle - \left\langle \hat{\mathcal{O}} \right\rangle \left\langle \hat{\mathcal{O}}^\prime \right\rangle \right)^2 + \left( \frac{1}{2i} \left\langle \left[ \hat{\mathcal{O}}, \hat{\mathcal{O}}^\prime \right] \right\rangle \right)^2} ~;~ \left\lbrace A, B \right\rbrace = AB +BA
\end{equation}

\end{itemize}
Let us denote the phase space operators by $\hat{Z}^c_\mu$ where $\mu = 1,2$ (see \eqref{M}) stands for our configuration space variables $\hat{T}_c, \hat{X}_c$ and $\mu = 3, 4$ refers to the conjugate momenta $\hat{P}_t, \hat{P}_x$. We write the first term in \eqref{sc_un_rel} as the square of the $\mu\nu$-th element of the variance matrix $V^0$ and the second term is identified as the square of the $\mu\nu$-th element of the symplectic matrix $\Omega^0$. Thus,
\begin{equation}
\Delta \hat{Z}^c_\mu \Delta \hat{Z}^c_\nu \geq \sqrt{\left( V^0_{\mu\nu} \right)^2 + \left( \Omega^0_{mu\nu} \right)^2}  ~;~  V^0_{\mu\nu} = \frac{1}{2} \langle \left\lbrace \hat{Z}^c_\mu, \hat{Z}^c_\nu \right\rbrace \rangle - \langle \hat{Z}^c_\mu \rangle \langle \hat{Z}^c_\nu \rangle ~\mathrm{and}~ \Omega^0_{\mu\nu} = \frac{1}{2i} \left[ \hat{Z}^c_\mu, \hat{Z}^c_\nu \right]
\end{equation}
Now as per Williamson's theorem \cite{Will} any arbitrary variance matrix $V^0$ can be diagonalised by a symplectic transformation $S \in Sp(2n, R)$ ($V^d = SV^0S^T ;~ \Omega^0 = S\Omega^0S^T$). The diagonalised $V^d$ will contain the symplectic eigenvalues of $V^0$, which can be shown to be at least doubly degenerate. $V^d$ will in general take a form $V^d = diag\left(\nu_1/2, \nu_2/2, \nu_1/2, \nu_2/2 \right)$. The symplectic spectrum, which is generically different from ordinary spectrum and can also be obtained as the ordinary spectrum of the object $\abs{2i\Omega^0 V^0}$ as a symplectic transformation of $V^0$ induces a similarity transformation in $\left( \Omega^0 V^0 \right)$ \cite{Zol}. In this diagonal form, $4$-dimensional phase space gets reduced to $2$-copies of independent $2$-dimensional phase space. In those $2$D spaces the uncertainty relation \eqref{sc_un_rel} can be written as,
\begin{equation}
\Delta \hat{Z}_\mu \Delta \hat{Z}_\nu \geq \sqrt{\left( V^0_{\mu\nu} \right)^2 + \left( \Omega^0_{\mu\nu} \right)^2} ~~~\mathrm{with}~\mu, \nu = 1,3 ~ (\mathrm{alternatively}~ 2,4) ~(\mathrm{no~sum~on~\mu,\nu})
\end{equation}
One can always choose, without loss of generality, the spread in $\hat{X}$ and $\hat{P}_x$ to be equal, thus equivalently $V^d_{11} = V^d_{33}$ and $V^d_{13} = V^d_{31} = 0$. Using the symplectic invariant form of $\Omega^0$ i.e. $\Omega^0_{11} = \Omega^0_{33} = 0$ and $\Omega^0_{13} = - \Omega^0_{31} = \frac{1}{2}$, one can finally arrive at a simpler form of the uncertainty relation in a separate $2$D copy of the total phase space
\begin{equation} \label{sc_un_rel_2}
det \, V^0 \geq \frac{1}{4}
\end{equation}
Thus, for the whole $4$-dimensional phase space, we will have
\begin{equation} \label{sc_un_rel_3}
det \, V^0 \geq \frac{1}{4^2}
\end{equation}
This gives us a symplectic $Sp(4, R)$ invariant form of the uncertainty relation. However, the present analysis was initially done at the level of commutative quantum mechanics and cannot be applied in a straightforward manner to a noncommutative quantum system, as Williamson's theorem may not be valid there. But under the  transformation \eqref{M} the variance matrices for commutative and noncommutative cases gets related as
\begin{equation}
V^0 = M V^\theta M^T ~\implies~ det(V^0) = det(V^\theta)
\end{equation}
indicating that here too we can write for the symplectic invariant form of uncertainty relation as
\begin{equation}
det \, V^\theta \geq \frac{1}{4^2}
\end{equation}



\end{document}